\documentclass[10pt]{article}
\usepackage{euscript,amssymb}

\font\msbm=msbm10
\def\RR{\hbox{\msbm R}}

\def\ZZ{\hbox{\msbm Z}}

\catcode`\@=11
\def\lesssim{\mathrel{\mathpalette\vereq<}}
\def\vereq#1#2{\lower3pt\vbox{\baselineskip1.5pt \lineskip1.5pt
\ialign{$\m@th#1\hfill##\hfil$\crcr#2\crcr\sim\crcr}}}
\def\gtrsim{\mathrel{\mathpalette\vereq>}}

\def\Let@{\relax\iffalse{\fi\let\\=\cr\iffalse}\fi}
\def\vspace@{\def\vspace##1{\crcr\noalign{\vskip##1\relax}}}
\def\multilimits@{\bgroup\vspace@\Let@
 \baselineskip\fontdimen10 \scriptfont\tw@
 \advance\baselineskip\fontdimen12 \scriptfont\tw@
 \lineskip\thr@@\fontdimen8 \scriptfont\thr@@
 \lineskiplimit\lineskip
 \vbox\bgroup\ialign\bgroup\hfil$\m@th\scriptstyle{##}$\hfil\crcr}
\def\Sb{_\multilimits@}
\def\endSb{\crcr\egroup\egroup\egroup}
\def\Sp{^\multilimits@}

\newcommand{\opensquare}{\mbox{$\rlap{$\sqcap$}\sqcup$}}
\newcommand{\be}[1]{\begin{equation}\label{#1}}
\newcommand{\ee}{\end{equation}}
\newcommand{\ba}[1]{\begin{eqnarray}\label{#1}}
\newcommand{\ea}{\end{eqnarray}}
\newcommand{\rf}[1]{(\ref{#1})}
\newcommand{\nn}{\nonumber}
\newcommand{\const}{\mbox{\rm const}}

\begin{document}
\author{Uwe G\"unther\dag\footnote{e-mail: u.guenther@htw-zittau.de} 
\ and
Alexander Zhuk\ddag\footnote{e-mail: zhuk@paco.odessa.ua}\\[2ex]
\dag Gravitationsprojekt, Mathematische Physik I,\\
Institut f\"ur Mathematik,
Universit\"at Potsdam,\\
Am Neuen Palais 10, PF 601553, D-14415 Potsdam, Germany\\[1ex]
\ddag
NASA/Fermilab Astrophysics Center,\\
Fermi National Accelerator Laboratory, MS 209, Box 500,\\
Batavia, Illinois 60510 - 0500\\[1ex]
and Department of Physics, University of Odessa, \\
2 Petra Velikogo St., Odessa 65100, Ukraine}
\title{A note on dynamical stabilization of internal spaces
in multidimensional cosmology}

\date{23.06.2000}

\maketitle

\abstract{The possibility of dynamical stabilization
of an internal space is investigated for
a multidimensional cosmological model with
 minimal coupled scalar field as inflaton.
It is shown that
a successful dynamical compactification crucially depends on
the type of interaction between the geometrical modulus field
and the inflaton and its decay products.
In the considered model a stable compactification can be ensured
via trapping
of the modulus field by a minimum of the effective potential.
}

\bigskip

\hspace*{0.950cm} PACS number(s): 04.50.+h, 98.80.Hw

%%%%%%%%%%%%%%%%%%%%%%%%%%%%%%%%%%%%%%%%%%%%%%%%%%%%%%%%%%%%

%%%%%%%%%%%%%%%%%%%%%%%%%%%%%%%%%%%%%%%%%%%%%%%%%%%%%%%%%%%%

\section{Introduction}
\setcounter{equation}{0}

\bigskip

It is well known that the fundamental physical constants in superstring
theories are related to the vacuum expectation values of the dilaton
and moduli fields, and variations of these fields would result
in variations of
the fundamental constants. In the context of standard Kaluza-Klein
models moduli are defined by the shape and size of the internal spaces
(we shall refer to the corresponding
fields as geometrical moduli). Up to now, there
are no experiments which show a variation of the fundamental constants
(theoretical and experimental bounds on such variations
can be found e.g. in \cite{1}). According to observations the internal spaces
should be static or nearly static at least from the time of recombination
(in some papers arguments are given in favor of the assumption that a
variation of the
fundamental constants is absent from the time of primordial
nucleosynthesis). Therefore, part of any realistic
multidimensional model should be  a mechanism for
moduli stabilization.

Within superstring theories such a stabilization is achieved, e.g., via
trapping of
the moduli fields by K\"ahler \cite{2} or racetrack \cite{3} potentials.
Recently it was pointed out in Ref. \cite{HSOW},
that in a cosmological setting
the trapping stabilization of
the dilaton field can  enhance
and become  robust due to the coupling of the dilaton
to the kinetic energy of ordinary matter fields.

Within  multidimensional cosmological models of
Kaluza-Klein type the problem of geometrical moduli stabilization
by effective potentials was subject of  numerous  investigations
\cite{4}.
It was shown that the scale factors of the internal spaces can
stabilize, e.g.,
in pure geometrical models  with
a bare cosmological constant and curved internal spaces as well as in
 models
with ordinary matter. Small conformal excitations of the internal space
metric near the minima of the effective potential have
the form of massive scalar fields  (gravitational excitons)
\cite{GZ(PRD)} developing in the external spacetime (later, since the
sub-millimeter weak-scale compactification approach these geometrical
moduli excitations are also known as radions).
Gravexitons were investigated
for a number of models in Refs. \cite{GZ}.  In Refs. \cite{4} - \cite{GZ}
the 4-dimensional Planck scale $M_{Pl(4)}$ was implicitly understood as 
the $D-$dimensional fundamental
scale\footnote{Hereafter, $D=D^{\prime} +4$ is the total
dimension of the multidimensional spacetime,  
$D^{\prime}$ --- the dimension of the internal components.}
and it was assumed that the internal spaces are 
compactified at sizes 
somewhere between the Planck scale $L_{Pl} \sim 10^{-33}$cm
and the Fermi scale
$L_{F} \sim 10^{-17}$cm to make them unobservable.

Recently it has been realized that the higher-dimensional
fundamental scale $M_{Pl(D)}$ can be lowered from the 4-dimensional
Planck scale $M_{Pl(4)}=1.22 \times 10^{19}$ GeV 
down to the Standard Model electroweak scale
$M_{Pl(4+D^{\prime})} \sim M_{EW}\sim 1$ TeV
providing by this way a new scenario for the resolution
of the hierarchy problem. The corresponding proposal 
led to an intensive study of various multidimensional theories.
The considered models can be roughly devided 
into two topological classes.

The first class consists of models with warped products of Einstein
spaces as internal spaces.  For simplicity, the corresponding
scale (warp) factors  are usually assumed as 
depending  only on
the coordinates of the external spacetime.
Whereas gravitational
interactions in such models can freely propagate
in all multidimensional (bulk) space, the Standard Model (SM)
matter is localized
on a 3-brane with thickness of order of the Fermi length
in the extra dimensions. 
Such a model was used in Ref. \cite{sub-mill1} for the demonstration of
the basic features of the
sub-millimeter weak-scale compactification hypothesis with
$M_{Pl(D)}\sim M_{EW}$.
Different aspects
of geometrical moduli stabilization for such models were considered
 e.g. in Refs. \cite{sub-mill}. A comparison of effective cosmological
 constant and gravexciton masses arising in the electroweak fundamental
 scale approach with those in a corresponding Planck scale approach was
 given in Ref. \cite{GZ(PRD2)}.

The second class consists of models following from  Ho\v rava-Witten
theory \cite{HW} where one starts from the strongly
coupled regime of $E_8\times E_8$ heterotic string theory and
interprets it as M-theory on an orbifold $\RR^{10}\times S^1/\ZZ_2$
with a set of $E_8$ gauge fields at each ten-dimensional orbifold
fixed plane.
After compactification on a Calabi-Yau three-fold and dimensional
reduction one arrives
at effective $5-$dimensional
solutions which describe a pair of parallel 3-branes
with opposite tension, and location at the orbifold planes.
For these models the $5-$dimensional metric contains 
a $4-$dimensional metric component  
 multiplied by a warp factor which is a  function of the additional
dimension \cite{RS}. The geometrical moduli stabilization for such models
was considered, e.g., in Refs. \cite{RS_stab}.
Difficulties of the stabilization in such models connected with the
required
unrealistic fine-tuning of the equation of state
on our 3-brane were pointed out in Ref. \cite{CF}, with possible resolutions
proposed in \cite{CGKT}.
Moreover, in Refs.
\cite{KKOP}
it was shown that the stabilization of the extra dimension
is a necessary
condition for the correct transition from  $5-$dimensional 
models with branes to the
standard $4-$dimensional Friedmann cosmology.
Various  other aspects
of cosmological brane world scenarios were investigated, e.g., in Refs.
 \cite{brane-cosmo}.

Usually,
the moduli stabilization is based on the trapping of
 the geometrical moduli fields at
a minimum of an effective potential so that the fields are static
(or may at most oscillate near
this minimum due to quantum fluctuations). However, in absence of any
minima, nothing forbids them
 to evolve very slowly as long as their evolution does not contradict
the observable data. It is clear that a
sufficiently slow evolution is allowed,
if the moduli fields came into this regime before
primordial nucleosynthesis.
Such an approach is in spirit of  Paul Steinhardt's
quintessence scenario \cite{quintessence}. It may happen that moduli fields
asymptotically tend to some limit. We shall call such a behavior dynamical
stabilization. An example for a dynamical stabilization
in modular cosmology
was presented  in Ref.
 \cite{Dine}. However, the interaction with ordinary
matter fields can destroy this stabilization mechanism.
 The situation occurs, e.g., in
the model considered in  sections \ref{setup}, and
\ref{e-frame} of the present paper.

Subject of the investigation in the present paper
is a multidimensional cosmological
model with a minimal coupled scalar field  as inflaton field.
In Ref. \cite{Maz} it was pointed out that a dynamical stabilization
of the geometrical modulus could be possible for
 a model with one Ricci-flat internal
space and a zero bulk cosmological constant. Here, we investigate this
model in more detail and show that the interaction of the geometrical modulus
field with the inflaton field in most cases
destroys the dynamical stabilization and leads
to  decompactification of the extra dimension. There are only two
 possible ways for
a stable compactification of the extra dimension.
For the first one, a dynamical stabilization,
one has to assume that the modulus field is only coupled to the inflaton
field but not to its decay
products. Due to the exponential decay of the inflaton during reheating
 the force term in the
modulus equation, obtained from the effective potential of the model,
decreases also exponentially. The present friction term
provides then the stabilization of the modulus field.
The second possibility for a stabilization
consists in the trapping of
the modulus field near a minimum of the effective potential.
As simple example we consider the effective potential from Ref.
\cite{GZ(PRD2)} where the minimum is generated by a non-Ricci-flat
 (curved) internal space and a non-zero bare cosmological constant.
The analysis does not depend on the
choice of the $D-$dimensional fundamental scale.

The paper is organized as follows. In section \ref{setup}
we explain the general
setup of our model and show a possible mechanism for a dynamical
stabilization of the geometrical modulus in zero-order approximation.
In section \ref{e-frame} we investigate the dynamical behavior of
the modulus and the inflaton field in more detail and show that the interaction
between them in most cases results
in a decompactification of the internal space. An example
for a stable compactification of the internal space by trapping of the
modulus field near the minimum of the effective potential is presented in
section \ref{stable-c}. The brief Conclusions of the paper (section \ref{conclusion})
are followed by an Appendix  on
higher dimensional perfect fluid potentials and
specific features of their dimensional reduction.

%%%%%%%%%%%%%%%%%%%%%%%%%%%%%%%%%%%%%%%%%%%%%%%%%%%%%%%%%%%%%%
\section{Model and general setup\label{setup}}
\setcounter{equation}{0}

\bigskip
We consider a cosmological model with metric
\begin{equation}
\label{2.1}g=g^{(0)}+e^{2\beta (x)}g^{(1)} \equiv
g^{(0)}+b^2(x) g^{(1)}\, ,
\end{equation}
which is defined on a manifold with warped 
product topology\footnote{This model can
be easily generalized to manifolds with warped product topology
of $n$ internal spaces:
$M=M_0\times M_1\times \dots \times M_n$
(see Refs. \cite{GZ(PRD),GZ(PRD2)}).}
\begin{equation}
\label{2.2}M=M_0\times M_1\, ,
\end{equation}
where $x$ are some coordinates of
the $D_0 =(d_0+1)$ - dimensional manifold $M_0$ and
\begin{equation}
\label{2.3}g^{(0)}=g_{\mu \nu }^{(0)}(x)dx^\mu \otimes dx^\nu .
\end{equation}
(The model corresponds to the type one class discussed in the Introduction.)

With  $D= D_0 +d_1$\ as total dimension, $\kappa_D ^2$ a
$D-$dimensional
gravitational constant and $S_{YGH}$ the standard York - Gibbons - Hawking
boundary term
%\cite{York,GH},
we choose the action functional in the form
\begin{equation}
\label{2.4}S=\frac 1{2\kappa_D ^2}\int\limits_Md^Dx\sqrt{|g|}
R[g]
-\frac12 \int\limits_Md^Dx\sqrt{|g|}\left( g^{MN}\partial_M \tilde \chi
\partial_N \tilde \chi + 2 U(\tilde \chi )\right) + S_{YGH}\, ,
\end{equation}
where the minimal coupled scalar field $\tilde \chi$ with an arbitrary
potential $U(\tilde \chi)$
depends on the external coordinates $x$ only.
This field can be understood as a zero mode of a bulk field.
{}From the other hand, such a scalar field can naturally
originate also in non-linear
$D-$dimensional theories \cite{GZ(non-lin)} where the metric ansatz \rf{2.1}
ensures its dependence on the $x$ coordinate only.

The main goal of the present paper consists in an investigation of the
internal space dynamical stabilization. As it was shown in \cite{Maz},
such a possibility exists for the considered model only if a bare
$D-$dimensional cosmological constant $\Lambda$ identically equals 
zero and the internal space is a Ricci-flat one:
$R\left[ g^{(1)}\right] = 0$.
%In contrast,
%for $\Lambda < 0$ and constant negative
%curvature internal space  $R\left[ g^{(1)}\right] = - d_1(d_1-1)$
%the internal space can be trapped by a minimum of an effective potential
%\cite{GZ(PRD2)}. We note that these two kinds of
%internal
%space stabilization are based on completely different mechanisms.

Let $b_0 = L_{Pl}e^{\beta_0}$ be the compactification scale of the
internal space at the present time and
\be{2.5}
v_{d_1}
\equiv v_0\times v_i \equiv b_0^{d_1} \times \int\limits_{M_1}d^{d_1}y
\sqrt{|g^{(1)}|}
\ee
the corresponding total volume of the internal space.
Instead of $\beta$ it is convenient to
introduce the shifted quantity:
\be{2.6}
\tilde \beta = \beta - \beta_0\, .
\ee
Then, after dimensional reduction action \rf{2.4} reads
\ba{2.7}
S&=&\frac 1{2\kappa _0^2}\int\limits_{M_0}d^{D_0}x\sqrt{|g^{(0)}|}%
e^{d_1\tilde \beta}\left\{ R\left[ g^{(0)}\right]
-d_1(1-d_1) g^{(0)\mu\nu }\partial _\mu \tilde \beta \,\partial _\nu
\tilde \beta - g^{(0)\mu \nu}\kappa^2_D
\partial_{\mu} \tilde \chi \partial_{\nu} \tilde \chi
-2\kappa^2_D U(\tilde \chi)\right\} \nn \\
&=& \int\limits_{M_0}d^{D_0}x\sqrt{|g^{(0)}|}
\left\{\Phi R\left[ g^{(0)}\right]
-\frac{1-d_1}{d_1} g^{(0)\mu\nu }\frac{\partial _\mu \Phi \,\partial _\nu
\Phi}{\Phi} -\kappa^2_0 \Phi
g^{(0)\mu \nu} \partial_{\mu} \chi \partial_{\nu} \chi
-2\kappa^2_0 \Phi V(\chi)\right\}\, ,
\ea
where
\be{2.8}
\Phi = \frac{1}{2\kappa^2_0}e^{d_1\tilde \beta}
=\frac{1}{2\kappa^2_0}\left(\frac{b}{b_0}\right)^{d_1}
\ee
and we redefined the scalar field $\tilde \chi$ and its potential as follows:
\be{2.8a}
\chi = \sqrt{v_{d_1}}\; \tilde \chi\; ,\quad V(\chi) =
v_{d_1}U\left(\chi /\sqrt{v_{d_1}}\right)\, .
\ee
In action \rf{2.7} $\kappa^2_0$
is the $D_0-$dimensional (4-dimensional) gravitational constant:
\be{2.9}
\kappa^2_0 = \frac{8\pi}{M^2_{Pl}}
:= \frac{\kappa^2_D}{v_{d_1}}\, ,
\ee
where $M_{Pl} = M_{Pl(4)}= 1.22\times 10^{19}\mbox{GeV}$. It is clear that the
scale of the internal space compactification $b_0$ is defined now by the
energetic scale of the $D-$dimensional gravitational constant
$\kappa^2_D$. If we normalize $\kappa^2_D$ in such a way that
$\kappa^2_D = 8\pi /M^{2+ d_1}_{EW}$, where
$M_{EW} \sim 1 \mbox{TeV}$ is the SM electroweak scale, then we
arrive at the intensively discussed  estimate
\cite{sub-mill1}:
$b_0 \sim 10^{(32/d_1) - 17} \mbox{cm}$. (Here, we assume for a toroidal
internal space $v_i\equiv 1$.) In the present paper we shall not
discuss different choices of $\kappa^2_D$. It has no influence on the main
conclusions about the possibility of an internal space
dynamical stabilization.
Generally speaking, we might not fix a scale of the internal
space stabilization in Eqs. \rf{2.6} - \rf{2.8}, but simply set
there $\beta_0 = 0$
and $b_0 = L_{Pl}$. For our concret model 
this does not mean that the stabilization takes place
at the planckian scale but results simply in a planckian normalization
of the $D-$dimensional gravitational constant: $\kappa^2_D \sim
M^{-(2+d_1)}_{Pl}$ \cite{GZ(PRD2)}.
In the latter approach, the scale of stabilization is
defined from the equations of motion.

To get the equations of motion in the Brans-Dicke frame it is convenient
to rewrite action \rf{2.7} as follows:
\be{1}
S = \int\limits_{M_0}d^{D_0}x\sqrt{|g^{(0)}|}
\left\{\Phi R\left[ g^{(0)}\right]
-\frac{\omega }{\Phi } g^{(0)\mu\nu }\partial _\mu \Phi \,\partial _\nu
\Phi + 2 \kappa^2_0 \Phi L_m \right\} \, ,
\ee
where $\omega := (1-d_1)/d_1 <0$ and
\be{2}
L_m = -\frac12 g^{(0)\mu \nu} \partial_{\mu} \chi \partial_{\nu} \chi
- V(\chi )\, .
\ee
Varying this action with respect to the metric $g^{(0)}_{\mu \nu}$
and the fields
$\Phi$ and $\chi$, we get respectively
\be{3}
R_{\mu \nu}-\frac12g^{(0)}_{\mu \nu}R = \kappa^2_0 T_{\mu \nu}+
\frac{\omega}{\Phi^2}\left[\Phi_{;\mu}\Phi_{;\nu}-\frac12g^{(0)}_{\mu \nu}
\Phi_{;\lambda}\Phi^{;\lambda}\right]
+\frac{1}{\Phi}\left[\Phi_{;\mu ;\nu}-g^{(0)}_{\mu \nu}
\opensquare \Phi \right]\, ,
\ee
\be{4}
R - \frac{\omega}{\Phi^2} g^{(0)\, \mu \nu}\partial_{\mu}\Phi
\partial_{\nu}\Phi + 2\kappa^2_0 L_m +\frac{2\omega}{\Phi} \opensquare \Phi
= 0\,
\ee
and
\be{5}
\opensquare \chi + \frac{1}{\Phi}  g^{(0)\, \mu \nu} \partial_{\mu} \Phi
\partial_{\nu} \chi = V^{\prime}\, ,
\ee
where
\ba{6}
T_{\mu \nu} &=& g^{(0)}_{\mu \nu} L_m 
+\partial_{\mu} \chi \partial_{\nu}\chi\, ,\nn\\
T&\equiv &T^{\mu}_{\mu} = D_0L_m +
g^{(0)\, \mu \nu} \partial_{\mu} \chi \partial_{\nu} \chi\, ,
\ea
\be{7}
\opensquare \equiv \frac{1}{\sqrt{|g^{(0)}|}} \partial_{\mu}\left(
\sqrt{|g^{(0)}|} g^{(0)\, \mu \nu} \partial_{\nu}\: \right)\, 
\ee
and the prime denotes the derivative with respect to the field $\chi$.

Contracting \rf{3} with respect to $g^{(0)\, \mu \nu}$ yields
\be{8}
R=\frac{\kappa^2_0}{(2-D_0)/2} T + \frac{\omega}{\Phi^2}
g^{(0)\, \mu \nu}\partial_{\mu}\Phi \partial_{\nu}\Phi
+\frac{1}{\Phi}\frac{D_0-1}{(D_0-2)/2} \opensquare \Phi\, .
\ee
Now, combining \rf{4} and \rf{8} we can rewrite the equation of motion for the
field $\Phi$ in the convenient form
\ba{9}
\frac{1}{\Phi}
\left(2\omega+\frac{D_0-1}{(D_0-2)/2}\right)\opensquare \Phi &=&
\frac{2\kappa^2_0}{D_0-2}T - 2\kappa^2_0 L_m \nn\\
&=& - \frac{4\kappa^2_0}{D_0-2}
V(\chi )\, .
\ea

Let us specify metric $g^{(0)}$ as
\be{10}
g^{(0)} = -dt\otimes dt + a^2(t)\bar g^{(0)}\, ,
\ee
where $\bar g^{(0)}$ is the metric of the $d_0-$dimensional Ricci-flat
external space: $R\left[\bar g^{(0)}\right] = kd_0(d_0-1) = 0\,
\Longrightarrow k=0 $ (in accordance with recent observations
our Universe is flat). For this metric we get
\be{11}
R_{00}[g^{(0)}] = -d_0\frac{\ddot a}{a} \quad \mbox{and} \quad
R[g^{(0)}] = d_0\left[2 \frac{\ddot a}{a} +(d_0 - 1)\frac{\dot a^2}{a^2}
\right]\, ,
\ee
where the dot
denotes the derivative with respect to the synchronous time $t$ in the
Brans-Dicke (BD) frame (string frame).

Below we consider homogeneous fields: $\Phi = \Phi (t)$ and $\chi = \chi(t)$.
Then, the $00-$component of the Einstein eq. \rf{3} reads
\be{12}
H^2 + \frac{2}{d_0-1}H\frac{\dot \Phi}{\Phi} = \frac{\omega}{d_0(d_0-1)}
\left(\frac{\dot \Phi}{\Phi}\right)^2 + \frac{2\kappa^2_0}{d_0(d_0-1)}
\left(\frac12 \dot \chi^2 + V\right)\, ,
\ee
where $H \equiv \dot a/a$ is the Hubble parameter.
If we take into account that for our metric \rf{10} and any homogeneous
field $\phi \,$ holds $\quad \opensquare \phi = -\ddot \phi -d_0 H \dot \phi$,
then we get for equations \rf{9} and \rf{5} correspondingly:
\be{13}
\ddot \Phi + d_0 H\dot \Phi = \frac{2\kappa^2_0}{\omega (d_0-1)
+ d_0} \Phi V(\chi )
\ee
and
\be{14}
\ddot \chi +d_0 H\dot \chi +\frac{\dot \Phi}{\Phi}\dot \chi = - V^{\prime}\, .
\ee

We suppose that the potential $V(\chi )$ of our model has a zero minimum:
$\left. V\right|_{min} = 0$.
%at the point $\chi = 0 \, :\, \left. V\right|_{\chi = 0}=0, \left. V^{\prime}
%\right|_{\chi =0}  = 0$.
Then, after the $\chi$ field has evoluted down-hill to this minimum 
and is frozen out, the right hand
side of eq. \rf{13} is equal to zero and the equation has a solution
\be{15}
\Phi (t) = \frac{1}{2\kappa^2_0}
-\dot \Phi_0 \int\limits_{t}^{\infty}
dt e^{-d_0\int\limits_{t_0}^{t} H dt}\, ,
\ee
where $\dot \Phi_0 := \dot \Phi (t = t_0)$ is an initial
value for $\dot \Phi$. The constant of integration in \rf{15}
is chosen in such a way that $\Phi (t \to \infty)
\longrightarrow 1/2\kappa^2_0$, i.e. that it corresponds to the stabilization 
of internal
space  $b \longrightarrow b_0$. Thus, in zero-order
approximation
(with respect to
the interaction between the $\Phi$ and $\chi$ fields in eq. \rf{13})
the dynamical stabilization of the internal space takes place
if the integral in eq. \rf{15} is convergent at the upper limit $t\to \infty$.
If the integral diverges the dynamical stabilization mechanism fails to work
and decompactification occurs.
A detailed analysis of this problem will be given in the next section
where we use a self-consistent approach and
investigate the influence of
the interaction between the fields on the dynamical stabilization
of the modulus field $\Phi$. It is more convenient to
perform the corresponding analysis in the Einstein frame. 
Obviously, if stabilization occurs
in the Einstein frame it has place also in the Brans-Dicke frame
and vice versa.

%%%%%%%%%%%%%%%%%%%%%%%%%%%%%%%%%%%%%%%%%%%%%%%%%%%%%%%%%%%%%
\section{The Einstein frame\label{e-frame}}
\setcounter{equation}{0}

\bigskip

A conformal transformation to the Einstein frame is given by equation
\begin{equation}
\label{2.10} g_{\mu \nu }^{(0)}= \Omega^2 \tilde g_{\mu \nu
}^{(0)} := {\left( e^{d_1\tilde \beta }\right) }^{-\frac{2}{D_0-2}}
\tilde g_{\mu \nu }^{(0)} = {\left(2\kappa^2_0
\Phi\right)}^{-\frac{2}{D_0-2}}
\tilde g_{\mu \nu }^{(0)}
\end{equation}
and yields  \cite{GZ(PRD)}
\begin{equation}
\label{2.11}S=\frac12\int\limits_{M_0}d^{D_0}x\sqrt{|\tilde
g^{(0)}|}\left\{ \frac{1}{\kappa^2_0} \tilde R\left[ \tilde g^{(0)}\right]
-\tilde g^{(0)\mu \nu }\partial _\mu \varphi \, \partial _\nu \varphi -
\tilde g^{(0)\mu \nu}
\partial_{\mu} \chi \, \partial_{\nu} \chi -
2U_{eff}(\varphi, \chi)\right\}\, ,
\end{equation}
where
\begin{equation}
\label{2.12}\varphi := \pm \frac{1}{\kappa_0} \sqrt{\frac{d_1(D-2)}{D_0-2}}
\tilde \beta
\end{equation}
and the effective potential can be written as
\begin{equation}
\label{2.13}
U_{eff}[\varphi , \chi ] := e^{2\kappa_0 \varphi
\sqrt{ \frac{d_1}{(D-2)(D_0-2)}}} V(\chi) :=
e^{2\sigma \kappa_0 \varphi} V(\chi)\, .
\end{equation}
(For definiteness we use the minus sign in eq. \rf{2.12}.)

The dimensionally reduced action \rf{2.11} describes a system of 4-dimensional
gravitational and scalar fields. The problem of the internal space
stabilization is reduced now to the investigation of the dynamics of these
fields. For this purpose  we specify the metric $\tilde g^{(0)}$.
In the Einstein frame the 4-dimensional metric \rf{10} can be rewritten
as 
\be{4.1}
\tilde g^{(0)} = - d\tilde t \otimes d\tilde t + \tilde a^2(\tilde t)
\bar g^{(0)} \, ,
\ee
where 'tilded' quantities are related to the Einstein frame.

Then, the equations of motion following from action \rf{2.11} read:
\be{4.2}
\tilde H^2 \equiv \left( \frac{\dot{\tilde a}}{\tilde a}\right)^2 =
\frac{2\kappa^2_0}{d_0(d_0-1)}\left\{\frac12\dot \varphi^2 +\frac12\dot
\chi^2 +  U_{eff}\right\}\, ,
\ee
\be{4.3}
\ddot \varphi + d_0 \tilde H \dot \varphi = -\frac{\partial U_{eff}}
{\partial \varphi} = -2\kappa_0 \sigma e^{2\sigma \kappa_0 \varphi}
V(\chi )\, ,
\ee
\be{4.4}
\ddot \chi + d_0 \tilde H \dot \chi = -\frac{\partial U_{eff}}
{\partial \chi} = - e^{2\sigma \kappa_0 \varphi} V^{\prime}(\chi )\, .
\ee
Hereafter, dots denote derivatives with respect to the
synchronous time $\tilde t$ in the Einstein frame, and primes
denote derivatives
with respect to the inflaton field $\chi $.
Further on, we assume for the external space dimension the usual value
$d_0 =3$. For this choice of $d_0$
the parameter $\sigma $ is fixed by the dimension of the extra space as
$\sigma = \sqrt{d_1/2(d_1+2)}$.

Inflation in our particular model \rf{4.2} - \rf{4.4} is known as soft
inflation and was extensively studied in Refs. \cite{BM,Maz}.
Here we simply quote the corresponding results.
During inflation the slow roll conditions $\dot \varphi ^2 \ll 2U_{eff}$,
$\dot \chi ^2 \ll 2U_{eff}$, $\ddot \varphi \ll 3\tilde H\dot \varphi $ and
$\ddot \chi \ll 3\tilde H\dot \chi $ hold and the Eqs. \rf{4.2} - \rf{4.4}
reduce to the simpler form
\ba{4.5}
3\tilde H\dot \varphi
&\approx &-2\sigma \kappa_0 e^{2\sigma \kappa_0\varphi}V(\chi)\, ,\nn\\
3\tilde H\dot \chi
&\approx &- e^{2\sigma \kappa_0\varphi}V^{\prime}(\chi)\, ,\nn\\
\tilde H^2
&\approx &\frac{\kappa _0^2}{3}U_{eff}
= \frac{\kappa _0^2}{3}e^{2\sigma \kappa_0\varphi}V(\chi)\, .
\ea
As result, the fields $\chi$ and $\varphi$ are connected with
the scale factor of the external space by the relations
\ba{4.11}
&&\dot \varphi \approx -\frac{2\sigma}{\kappa_0}\tilde H \nn\\
\Longrightarrow &&
\kappa_0 (\varphi -\varphi_0)\approx -2\sigma \int \tilde H d\tilde t 
=-2\sigma \ln \frac{\tilde a}{\tilde a_0}
\ea
and
\be{4.11a}
\kappa_0^2\int ^{\chi}_{\chi_0}d\chi \frac{V}{V^{\prime}}=
\ln \frac{\tilde a}{\tilde a_0}\, ,
\ee
where the initial conditions 
are denoted by the subscript "0" and
are fixed at $\tilde t_0$, when the Universe enters the inflationary phase.
{}From \rf{2.12}, \rf{4.11} we see that during the inflationary stage of
the external spacetime the scale factor of the
internal space undergoes inflation too (see e.g. also \cite{Maz})
\be{4.12}
b(\tilde t) = \bar b_0 e^{\frac{2}{D-2}\int \tilde H d\tilde t}\, .
\ee
The necessity of such a stage was stressed in paper \cite{KL},
 where it was shown
that to solve the horizon and flatness problem, there must be a stage of
inflation in the bulk space before the compactification of the internal spaces
can be completed.
If we take into account that during this stage $\tilde a \sim
\exp{(\int \tilde H d\tilde t)}$
then we get following relation between initial and final
values: $b_f/b_i = [a_f/a_i]^{2/(D-2)}$ (see also \cite{Maz}).

Let us investigate now the dynamical behavior of the model at the stage when
the inflaton scalar field $\chi$ is evoluting down-hill to the minimum of its
effective potential and is located in the very vicinity of this
 minimum.
Without loss of generality,
we suppose that potential $V(\chi )$ has a zero minimum at the point
$\chi = 0 \, :\, \left. V\right|_{\chi = 0}=0, \left. V^{\prime}
\right|_{\chi =0}  = 0\, , \left. V^{\prime \prime}\right|_{\chi =0}  > 0$.
This leads to two implications.

First, for $|\chi |< 1$ (i.e. $|\chi |<
M_{Pl}$) this potential can be crudely approximated by
\be{4.13}
V(\chi ) \approx \frac12 m^2_{\chi}\chi^2\, ,
\ee
where $m^2_{\chi} \equiv \left. V^{\prime \prime}\right|_{\chi = 0}$.
In the same approximation we get for
eqs. \rf{4.3} and \rf{4.4} correspondingly:
\be{4.14}
\ddot \varphi + 3 \tilde H \dot \varphi + \kappa_0 \sigma
e^{2\sigma \kappa_0 \varphi} m^2_{\chi}\chi^2 = 0
\ee
and
\be{4.15}
\ddot \chi + 3 \tilde H \dot \chi + e^{2\sigma \kappa_0 \varphi}
m^2_{\chi} \, \chi = 0\, .
\ee

Second, from the
structure of the effective potential 
$U_{eff}(\chi ,\varphi )=e^{2\sigma \kappa _0\varphi}V(\chi )$
it is clear that the zero minimum 
$\left. V\right|_{\chi = 0}=0$ is globally  degenerate and, in crude analogy
with Goldstone bosons in $\varphi^4 -$theories, the
field $\varphi $ plays the role of the 
zero mode along the degeneration
line $\chi =0$ in the $(\chi, \varphi)-$plane.
This is easy to see from the Hessian
of the effective potential
\be{4.15a}
\left(
\begin{array}{cc}
\partial ^2_{\varphi \varphi}&\partial ^2_{\varphi \chi}\\
\partial ^2_{\chi \varphi}&\partial ^2_{\chi \chi}
\end{array}
\right)
\left.U_{eff}\right|_{\chi = 0}
=\left(
\begin{array}{cc}
0 &0\\
0&e^{2\sigma \kappa _0\varphi}m^2_{\chi}
\end{array}
\right)\, ,
\ee
which defines the mass matrix of the normal modes of the model at the point
$\chi = 0$ (see also Ref. \cite{GZ(PRD)}). This means that the geometrical
modulus field $\varphi $
corresponds to a flat direction of the effective
potential $U_{eff}$ and is not stabilized by a minimum of this potential (see
 e.g. Ref. \cite{FS}).
Due to the remaining $\varphi -$dependence of the effective mass of the 
$\chi -$mode the analogy with Goldstone bosons in $\varphi ^4-$theories 
is only very crude.

As first step of our analysis,
we investigate the equation of motion \rf{4.14}
of the modulus field not taking into account the energy constraint \rf{4.2}
and
assuming that the $\chi$ field is frozen in the minimum of the potential
$V(\chi )$ (zero-order approximation in the fluctuations of field $\chi$):
\be{4.16}
\chi  \longrightarrow 0 \quad \Longrightarrow \quad
\ddot \varphi + 3\tilde H \dot \varphi \approx 0\, .
\ee
For a cosmologically non-damped evolution with $\tilde H=0$ we would
have $\ddot \varphi \approx 0$ and a dynamically non-stabilized behavior of the
modulus field
$\varphi =\dot \varphi _0 \tilde t +\varphi _0$, 
where the initial values are chosen 
at some $\tilde t=\tilde t_0\,:\quad
\dot \varphi _0 :=\dot \varphi (\tilde t=\tilde t_0),\quad
\varphi _0 := \varphi (\tilde t=\tilde t_0)$.
Due to the damping term $3\tilde H\dot \varphi \neq 0$ eq. \rf{4.16}
 has a solution
\be{4.17}
\varphi (\tilde t) = -\dot \varphi_0 \int\limits_{\tilde t}^{\infty}
d\tilde t e^{-3\int\limits_{\tilde t_0}^{\tilde t}\tilde H d\tilde t}\, ,
\ee
which describes a dynamical (asymptotical) stabilization if the integral 
converges at its upper limit $\tilde t\to \infty $.
The constant of integration in \rf{4.17}
is chosen in such a way that $\varphi (\tilde t \to \infty)
\longrightarrow 0$ and  corresponds to the internal space stabilization at
$b \longrightarrow b_0$.
Obviously, this solution is the Einstein frame analogue of
the Brans-Dicke frame solution
\rf{15} and reproduces the dynamical stabilization scenario of Ref. \cite{Maz}.
For a power-law behavior of the external scale factor:
$\tilde a \sim \tilde t^s \Longrightarrow
\tilde H = \dot{\tilde a} / \tilde a = s
/
\tilde t$ ($s = 1/2\, , 2/3 $ corresponds to the radiation dominated (RD)
and matter dominated (MD) era respectively) this equation yields
\be{4.18}
\varphi (\tilde t) = - \frac{\dot \varphi_0}{3s-1}\frac{\tilde t_0^{\, 3s}}
{\tilde t^{\, 3s-1}} \, , \qquad s > \frac13 \, .
\ee
We can use this equation for estimates of the variation of the fundamental
constants due to the dynamics of the internal space. Usually \cite{1},
such variations are proportional to $\dot b / b$. 
{}From eq. \rf{4.18}
we get
\be{4.19}
\frac{\dot b}{b} = - \sqrt{\frac{D_0-2}{d_1(D-2)}}\kappa_0\dot \varphi_0
\left(\frac{\tilde t_0}{\tilde t}\right)^{3s}\, .
\ee
We choose the initial values $\dot \varphi_0$ and $\tilde t_0$
 in such way that they correspond to the end of inflation:
$\kappa_0 \dot \varphi_0 = - 2\sigma \tilde H_e$ (see eq. \rf{4.11})
and $\tilde H_e \tilde t_0 \sim 70$. Then, eq. \rf{4.19} yields
\be{4.20}
\frac{\dot b}{b} \approx \frac{2}{D-2} \tilde H_e
\left(\frac{70}{\tilde H_e \tilde t}\right)^{3s}\, .
\ee
For the simplest models of inflation, as in our case, the COBE data
predict $\tilde H_e \sim 10^{-5} M_{Pl}$.
Thus, we obtain $\left.\dot b / b \right|_{\tilde t \sim 10^3
sec\, ; s =1/2} \sim 10^{-14}\mbox{yr}^{-1}$. This means that effectively
there is no variation
of the fundamental constants starting from the time of
nucleosynthesis. For the modulus field value we can easily get
an estimate:
\be{4.20a} \kappa_0 \varphi (\tilde t) \approx \frac{140 \sigma }{3s-1}
\left(70/\tilde H_e\tilde t\right)^{3s-1} \Longrightarrow
\left. \kappa_0 \varphi \right|_{\tilde t \sim 10^3
sec\, ; s =1/2} \sim 10^{-18} \ll 1\, .
\ee
Thus, in zero order approximation
and not taking account of the constraint \rf{4.2},
the internal space would appearently  stabilize to this time.

Let us now use a self-consistent approach to our model
taking also into account the constraint  \rf{4.2}.
For frozen
$\chi$ field we have $\chi =\chi_0=0$, $\dot \chi _0=0$ and $V(\chi_0)=0$
so that the energy density
$\rho_{\chi (E)}\equiv \frac12\dot\chi ^2+U_{eff}$ vanishes
($\rho_{\chi (E)}(\chi _0)=0$) and the 
 equation system \rf{4.2} - \rf{4.4} reads:
\begin{eqnarray}
\tilde H^2 &=&\frac{\kappa _0^2}{6}\dot \varphi ^2\, ,\label{4.18a}\\
\ddot \varphi +3\tilde H\dot \varphi &=&0\, .\label{4.18b}
\end{eqnarray}
The solutions are easily found as
\be{4.18c}
\tilde H=\frac{1}{3\tilde t},\quad \tilde a \sim \tilde t^{1/3},\quad
\kappa _0\dot \varphi =\pm \sqrt{\frac{2}{3}}\frac{1}{\tilde t},
\quad \kappa _0\varphi 
=\pm \sqrt{\frac{2}{3}}\ln {\frac{\tilde t}{\tilde t_0}} +\kappa _0\varphi _0
\ee
and show that the possible dynamical stabilization mechanism,
which holds for $s>1/3$, fails to work in the case $s=1/3$
due to the logarithmically diverging $\varphi$.
Heuristically, we can say that for frozen inflaton field $\chi$
the decoupled modulus field $\varphi $
 behaves like an ultra-stiff perfect fluid\footnote{See \cite{Z(CQG),SWZ}
 and the
Appendix for a discussion of the equivalence between a scalar field
$\varphi$ and a perfect fluid with energy density $\rho _{\varphi}$, pressure
$P_{\varphi}$ and equation of state
$P_{\varphi}=(\alpha -1)\rho_{\varphi}$.}
with
equation of state $P_{\varphi (E)}=\rho_{\varphi (E)}\sim \tilde a^{-6}$
and produces not enough "cosmological friction"
$\tilde H=s/\tilde t$  in order to
come to a rest at some finite value $|\varphi (\tilde t\to\infty)|<\infty$.
As consequence the additional
dimensions cannot stabilize, but rather they decompactify. 

This behavior can be partially circumvented by passing from the homogeneous 
modulus approach $\varphi =\varphi (\tilde t)$ to an inhomogeneous one
$\varphi =\varphi (\tilde x)$ with modulus fluctuations behaving like 
radiation (see e.g. \cite{Dine,BBMSS}). 
The corresponding energy density could lead
to the needed "cosmological friction" $\tilde H$ with $s>1/3$.
But as we will show below in this section, the interaction of the modulus 
field with the inflaton will destroy such a dynamical stabilization mechanism.
For simplicity, we will restrict our subsequent considerations 
to purely homogeneous fields.

For such fields we can circumvent the decompactification mechanism \rf{4.18c} 
if we assume that $\chi$ performs
small fluctuations around the minimum of its potential $V(\chi)$ yielding 
by this way a nonvanishing energy density $\rho_{\chi (E)}>0$ 
which could provide the
needed "cosmological friction" $\tilde H$ with $s>1/3$.
So, we shall investigate
corrections to the  equation system \rf{4.2},  \rf{4.14}, \rf{4.15}
due to the
dynamics of the
field $\chi $.
To achieve this goal, we assume that the modulus field $\varphi$ is already
nearly
stabilized at $\varphi \approx 0$ and
 embed our system \rf{4.14}, \rf{4.15}  in a generalized 
astrophysical setting allowing for decay processes
of the inflaton field $\chi$ into  usual matter.
The corresponding particle interactions
result in a polarization operator $\Pi$
of the field $\chi$ which shifts the squared mass
in eq. \rf{4.15}: $m^2_{\chi}\longrightarrow m^2_{\chi}+\Pi$
(see e.g. \cite{KLS}).
The imaginary part of $\Pi$ is responsible for the inflaton decay
and is connected with the decay rate $\Gamma _{\chi}$
by the relation Im$\Pi =m_{\chi}\Gamma _{\chi}$.
It is shown in Ref. \cite{KLS} that phenomenologically such a decay can be
taken into account by adding an extra friction
term $\Gamma \dot \chi$ to the classical equation of motion \rf{4.15}
(instead of adding the term proportional to the imaginary part
of the polarization operator):
\be{4.20b}
\ddot \chi +(3\tilde H+\Gamma _{\chi})\dot \chi
+e^{2\sigma \kappa _0 \varphi } m^2_{\chi}\chi =0\, .
\ee
We use this equation to crudely understand the dynamical behavior
of the field $\chi $ during the post-inflationary evolution
of the Universe (not taking into account preheating or subtleties
of the decay processes \cite{KLS}).
 It is convenient to investigate 
eq. \rf{4.20b} with the help of a substitution
(see \cite{Nayfeh}, Chapter 14):
\be{4.21}
\chi (\tilde t) := B(\tilde t) u(\tilde t) :=
e^{-\frac12(\Gamma _{\chi}\tilde t+3
\int \tilde H d\tilde t)} u(\tilde t)\, ,
\ee
where the function $u(\tilde t )$ satisfies the equation
\be{4.22}
\ddot u + \left[ \bar m^2_{\chi} - \frac14 (3\tilde H+\Gamma _{\chi})^2
- \frac32 \dot
{\tilde H}\right] u = 0
\ee
and $\bar m^2_{\chi}(\tilde t) = m^2_{\chi} \exp\, (2\sigma \kappa_0
\varphi)$.
If we suppose further that
$(3\tilde H+\Gamma_{\chi})^2, \dot{\tilde H},
(\dot{\bar m_{\chi}}/\bar m_{\chi})^2$ are
small
compared with $\bar m^2_{\chi}$ (which, for a viable model should
take place in the large $\tilde t \gg t_0$ limit when
$|\kappa_0 \varphi |\ll 1$), then eq.
\rf{4.22} has an approximate solution of the form (see also \cite{PWW})
\be{4.23}
u(\tilde t) \approx \cos (\bar m_{\chi} \tilde t)
\quad \Longrightarrow \quad \chi (\tilde t) \approx B(\tilde t)
\cos (\bar m_{\chi} \tilde t) \, .
\ee
It can be easily seen from the definition of $B(\tilde t)$ that this
function satisfies the equation 
\be{4.24}
\frac{d}{d\tilde t}\left( \tilde a^3B^2 \right)
= -\Gamma _{\chi} \tilde a^3B^2 \, .
\ee
Approximating the energy density of the inflaton and the corresponding 
number density as 
\be{4.25}
\rho_{\chi (E)} = \frac12 \dot \chi^2 
+ \frac12 \bar m^2_{\chi}\, \chi^2 \approx
\frac12 B^2 \bar m^2_{\chi}\, ,
\quad n_{\chi (E)}\approx \frac12 B^2 \bar m_{\chi}
\ee
shows that they satisfy the differential relations
\be{4.26}
\frac{d}{d\tilde t}(\tilde a^3\rho _{\chi (E)})
=-\Gamma _{\chi}\tilde a^3\rho _{\chi (E)}
\qquad \mbox{\rm and} \qquad
\frac{d}{d\tilde t}(\tilde a^3n_{\chi (E)})
=-\Gamma _{\chi}\tilde a^3n_{\chi (E)}
\ee
with solutions
$\rho _{\chi (E)} \sim e^{-\Gamma _{\chi }\tilde t}\tilde a^{-3}$
and
$n_{\chi (E)}\sim e^{-\Gamma _{\chi}\tilde t}\tilde a^{-3}$.
{}From these relations we see  that during the stage
$\bar m_{\chi}>\tilde H \gg \Gamma _{\chi}$
the inflaton performs damped oscillations with an energy density
corresponding to a dust-like
perfect fluid $(\rho _{\chi (E)}\sim \tilde a^{-3})$ with slow decay
$\sim e^{-\Gamma _{\chi}\tilde t}\sim 1$.
This reheating stage ends when $\tilde H\lesssim \Gamma _{\chi}$ and
the evolution of the energy density is dominated
by the exponential decrease due to decay with rate $\Gamma _{\chi}$.

If the decay rate $\Gamma_{\chi }$ of the inflaton particles into
usual Standard Matter (SM)
is sufficiently large,
then starting from the
characteristic decay time $\tilde t_D \sim
\Gamma^{-1}_{\chi}$, i.e. from the time of the most intensive reheating,
the energy density of the corresponding
relativistic particles behaves as
$\rho_{SM (E)} \sim \tilde a^{-4}$.
Clearly, the energy loss of the $\chi$ field due to the decay process is
accompanied by a corresponding energy increase of the decay products. 
As result,
the effective energy density 
$\rho _{eff (E)}:=\rho _{\chi (E)}+\rho _{SM (E)}$ 
is only deluted by the cosmological
expansion and can be roughly
approximated as
\be{4.26aa}
\rho _{eff (E)}=\rho _{\chi (E)}+\rho _{SM (E)}\sim
e^{-\Gamma _{\chi}\tilde t}\tilde a^{-3}
+(e^{-\Gamma _{\chi}\tilde t_1}-e^{-\Gamma _{\chi}\tilde t})\tilde a^{-4}\sim
\tilde a^{-3}, \tilde a^{-4}\, .
\ee
This relation holds for times
$\tilde t \gtrsim \tilde t_1 \sim \bar m_{\chi}^{-1}$
with $\Gamma _{\chi}\tilde t_1\ll 1$,
where $\tilde t_1$  plays the role of an effective initial time which fixes
the begin of the coherent $\chi$ oscillations and 
of the decay process.
{}From the Einstein equations of the extended interacting system we can
derive a corresponding Friedman equation similar to \rf{4.2}. 
Obviously, for a viable model the internal
space should be stabilized before the nucleosynthesis, i.e.
inequality $|\kappa^2_0 \dot \varphi |\ll 1$ should take place before this
stage. Then, starting from this moment,
the Hubble parameter is defined from this extended Friedman equation
by $\rho_{eff (E)}$:
$\tilde H^2 \approx \frac13 \kappa_0^2 \rho_{eff (E)}$.
Thus, the Hubble parameter during the post-inflationary stage,
including the period of nucleosynthesis, is defined
by the energy densities $\rho _{\chi (E)}$ and $\rho _{SM (E)}$ and 
 can be roughly approximated as
$\tilde H
= s/\tilde t$ where $s=1/2\, , 2/3$ for $\rho_{eff (E)} \sim \tilde a^{-4}\, ,
\tilde a^{-3}$ respectively\footnote{We note that the expansion 
parameter $s$ is connected with the parameter $\alpha $ 
in the equation of state $P_{eff (E)}=(\alpha -1)\rho _{eff (E)}$ 
by the relation $s=\frac{2}{d_0\alpha}$. See also \rf{a15} - \rf{a17}.}. 
Hence, in the same approximation we have
\be{4.26a}\rho _{eff (E)}\approx \frac{3s^2}{\kappa ^2_0}\tilde t^{-2}\, .
\ee

Let us now in some detail consider the required asymptotic evolution 
of the modulus field $\varphi \approx 0$
during the post-inflationary stage up to the period of nucleosynthesis,
when the modulus stabilization should be finished.
{}From eq. \rf{a13} given in the Appendix we see
that during the inflaton dominated post-inflationary
stage the equation of motion \rf{4.14} for the modulus field $\varphi$
can be rewritten as:
\be{4.27}
\ddot \varphi + 3 \tilde H \dot \varphi
=(\alpha_{\chi} -2)\sigma \kappa_0  \rho_{\chi (E)} \, .
\ee
After this stage, the further evolutional behavior of the modulus field
crucially depends on its coupling to the decay products
of the inflaton field, i.e. on the coupling type to the SM fields.
We will illustrate this specific model dependent feature
with the help of the two most simplest examples,
a model (I) where the decay products of the inflaton
have a similar coupling to the modulus field like  the inflaton itself,
and a model (II) where the decay products are
not coupled to the modulus field at all.

Model (I):  Under a model with
similar coupling of the inflaton field and its decay
products (the SM fields)
to the modulus field we understand a model where the evolution
of the modulus field is defined via extension of \rf{4.3}
by an equation of motion of the type
\ba{4.27a}
\ddot \varphi + 3 \tilde H \dot \varphi
&=&(\alpha_{\chi} -2)\sigma\kappa_0
\rho_{\chi (E)}+(\alpha_{SM} -2)\sigma\kappa_0\rho_{SM (E)} \nn\\
&\approx & (\alpha _{eff} -2)\sigma\kappa _0\rho_{eff (E)}\, .
\ea
Let us further assume that the required asymptotic stabilization 
of the modulus field is almost achieved
$ \varphi \approx 0$
so that  eq. \rf{4.27a} can be considered as a
non-homogeneous differential equation. 
In this approximation the general solution of \rf{4.27a}
can be written as a sum of the general solution of the
homogeneous eq. \rf{4.16} and a particular solution of the
non-homogeneous equation. As a solution of the homogeneous equation we
can take eq. \rf{4.18}. Then, using \rf{4.26a} and \rf{a17} 
the solution of eq. \rf{4.27a} can be approximated as
\be{4.28}
\varphi (\tilde t) \approx
- \frac{\dot \varphi_0}{3s-1}\frac{\tilde t_0^{\, 3s}}
{\tilde t^{\, 3s-1}} -
\frac{2s\sigma }{\kappa_0}\ln \frac{\tilde t}{T_0}
\ee
with $s=1/2, 2/3$ for $\rho _{eff}\sim a^{-4}, a^{-3}$ respectively.
The time  $T_0$ plays to role of an effective initial time
which is defined from the value of the
internal scale factor $b(\tilde t)$ at the beginning of the evolutional stage
described by \rf{4.27a}.
{}From eq. \rf{4.28} we see that the used perfect fluid ansatz leads to
a decompactification of the internal space. For times $\tilde t \gg \tilde
t_0$,
the internal scale factor behaves as $b \approx b_0\left( \tilde t/T_0
\right)^{\gamma}$, where $\gamma = 2s/(D-2)$ and e.g.
$\gamma = 1/4\, , 1/3$ for $d_1 =2$ and $s = 1/2\, ,2/3$ respectively.
Eq. \rf{4.28} shows also that for $\tilde t \gg t_0$ we obtain
$\dot b/b \sim -\kappa_0 \dot \varphi \sim \tilde t^{-1}$, so that
via extrapolation to our present time we would get the estimate
$10^{-10}\mbox{yr}^{-1}$. This estimate is
much greater than $10^{-14}\mbox{yr}^{-1} $ following from observations.
So, solution \rf{4.28} leads to a decompactification of
the internal space  and
does not ensure the necessary stabilization
of the internal spaces at the present time.
This was proved by the rule of contraries. First, we supposed that
the modulus stabilization occurs and we can use the approximation
 \rf{4.26a}.
Then, we showed that our proposal is
wrong because an unboundedly increasing destabilization 
term occurs in the solution \rf{4.28}.

Let us now consider model (II) with vanishing coupling of the decay products 
of the inflaton to the modulus field\footnote{But
the decay products still define
the dynamical behavior of the Universe.}.
In this case eq. \rf{4.27} of the modulus evolution
holds also after the inflaton-dominated era. Using the approximation
\rf{4.26a}
for the energy density $\rho _{\chi}$ of the
decaying dust-like perfect fluid component we rewrite this equation  with
an ansatz analogous to \rf{4.26a} and $s=2/3$ as
\be{4.28a}
\ddot \varphi +2\tilde t^{-1} \dot \varphi
= - \frac{4\sigma}{3\kappa_0}\tilde t^{-2}e^{-\Gamma _{\chi}\tilde t}\, .
\ee
For simplicity of notations we introduce the abbreviation
$q:=4\sigma /(3\kappa _0\Gamma _{\chi})$.
Then, the solution of \rf{4.28a} with internal space stabilization
$\varphi (\tilde t\to \infty) \longrightarrow 0$ at $b\longrightarrow b_0$
and initial condition 
$\dot \varphi (\tilde t =\tilde t_0)=\dot \varphi _0$ 
can be easily found as
\be{4.28b}
\varphi (\tilde t) =
\left(qe^{-\Gamma _{\chi } \tilde t_0}-\dot \varphi _0 \tilde t^2_0 
\right)
\frac{1}{\tilde t}
-q\int ^{\infty}_{\tilde t} \frac{e^{-\Gamma _{\chi }t}}{t^2}\, dt \, .
\ee
The corresponding characteristic variation of the internal scale factor
for the same initial conditions on $\tilde t_0$ and $\dot \varphi _0$ as
for the estimate \rf{4.20}
reads
\be{4.28c}
\frac{\dot b}{b}=-\kappa_0\sqrt{\frac{D_0-2}{d_1(D-2)}}
\left[q\left(e^{-\Gamma _{\chi}\tilde t}
-e^{-\Gamma _{\chi}\tilde t_0}\right)+
\dot \varphi _0 \tilde t^2_0\right]\frac{1}{\tilde t^2}
\ee
and gives for a decay channel of the inflaton to fermions with
decay rate \cite{infl} $\Gamma _{\chi}\sim 10^{-12}M_{Pl}$ the estimate
$\left.\dot b/b\right|_{\tilde t\sim 10^3 sec,d_1=2}\sim
10^{-31} yr^{-1}$. The modulus field stabilizes at the same time at
$\left.\kappa _0\varphi\right|_{\tilde t \sim 10^3 sec,d_1=2}\sim
10^{-35}\ll 1$.
Thus, there exists a possible dynamical
stabilization scenario for a decaying
inflaton field and a modulus field which is not coupled to the decay products.
Clearly, 
this dynamical stabilization scenario is rather artificial and in general
the decay products will be also functionally coupled to the modulus field 
what can destroy the stabilization.

Above, we considered a model with zero effective cosmological constant.
However, resent observations show the existence of a positive cosmological
constant $\Lambda \sim 10^{-57}\mbox{cm}^{-2}$ for our Universe.
So, it is of interest
to include such a $\Lambda-$term into our consideration.
This can be easily done
if we suppose that the potential $V(\chi )$ has a non-zero minimum at
$\chi = 0$. Thus, for $|\chi |\lesssim M_{Pl}$ the potential reads
\be{4.29}
V(\chi ) \approx V_0 + \frac12 m^2_{\chi}\chi^2\, .
\ee

It is clear that $\Lambda_{eff} \equiv e^{2\sigma \kappa_0 \varphi}V_0$
plays the role of the effective cosmological constant which asymptotically
tends to $V_0$ in the case of the internal space stabilization $\varphi
\longrightarrow 0$.
Such a behavior of the effective cosmological constant is similar to
the quintessence scenario \cite{quintessence}.
We suppose that in accordance with observations
$\kappa_0^2 V_0 \sim 10^{-57}\mbox{cm}^{-2} > 0$.
For such values of $V_0$
the influence of $V_0$ on the Universe and the fields dynamics
becomes essential only
at a stage close to our present time. During earlier post-inflationary
evolution stages
$e^{2\sigma \kappa_0 \varphi}V_0$
is negligible compared with $\rho_{\chi (E)}$:
\  $e^{2\sigma \kappa_0 \varphi}V_0 \ll \rho_{\chi (E)}
= \frac12 \dot \chi^2
+ \frac12 \bar m^2_{\chi}\chi^2$.
But, with progression of the Universe expansion $\rho_{\chi (E)}$
decreases and
becomes less and less
compared with $e^{2\sigma \kappa_0 \varphi}V_0$.
Let us investigate the influence of $V_0$ on the
internal space stabilization when
$e^{2\sigma \kappa_0 \varphi}V_0 \gtrsim \rho_{\chi (E)}$. We also assume
that up to this time $|\kappa_0 \varphi |\ll 1$. Under this assumptions
eq. \rf{4.27} is modified as follows:
\be{4.30}
\ddot \varphi + 3 \tilde H \dot \varphi = -2\kappa_0 \sigma
e^{2\sigma \kappa_0 \varphi} V_0
+(\alpha -2)\kappa_0 \sigma \rho_{\chi (E)} \approx
-2\kappa_0 \sigma V_0\, .
\ee
A solution of this equation can be found similar to eq. \rf{4.28} and reads
\be{4.31}
\varphi (\tilde t) =
- \frac{\dot \varphi_0}{3s-1}\frac{\tilde t_0^{\, 3s}}
{\tilde t^{\, 3s-1}} -
\frac{\sigma \kappa_0 V_0}{3s+1} \tilde t^{\, 2}\, .
\ee
Thus, an effective cosmological constant
results also in a decompactification (destabilization) of the internal space
at late times.

Summarizing the results of the present section, we can conclude that the 
considered possible dynamical stabilization scenario is very sensitive
to the coupling of the modulus field to small fluctuations of the inflaton 
and matter fields, as well as to a non-vanishing vacuum contribution. Thus,
this scenario is in general not sufficiently robust and each concret model
needs a detailed study whether a dynamical modulus stabilization could work 
or not. In the next section we extend our toy model and induce 
a trapping mechanism to guarantee the modulus stabilization.  
%%%%%%%%%%%%%%%%%%%%%%%%%%%%%%%%%%%%%%%%%%%%%%%%%%%%%%%%%%
\section{Stable compactification\label{stable-c}}
\setcounter{equation}{0}

\bigskip
In this section we present an example for a model which ensures
stable compactification of the internal space by a trapping
of the geometrical modulus at the minimum of the effective potential.
For this purpose, we modify our setup model of section \ref{setup}
including
a non-zero bare $D-$dimensional cosmological constant $\Lambda$ into the
corresponding action functional \rf{2.4}, \rf{2.7} 
and assume that the internal space is not Ricci-flat
$R[g^{(1)}] \equiv R_1 = \const \ne 0$.
Then, instead of \rf{2.13} we get an effective potential
\ba{5.1}
U_{eff}[\varphi , \chi ] &=& e^{2\kappa_0\varphi \sqrt{ \frac{d_1}
{(D-2)(D_0-2)}}}\left[ -\frac{1}{2\kappa^2_0}
\tilde R_1e^{2\kappa_0\varphi \sqrt{ \frac{D_0-2}{d_1(D-2)}}}
+\frac{1}{\kappa^2_0}\Lambda + V(\chi)\right]\equiv \nn \\
&\equiv & e^{2\sigma \kappa_0 \varphi}
\left[ -\frac{1}{2\kappa_0^2}\tilde R_1e^{2\gamma \kappa_0 \varphi}
+\frac{1}{\kappa_0^2}\Lambda + V(\chi)\right] \, ,
\ea
where $\tilde R_1 := R_1/b^2_0$ and we used the obvious abbreviations
$\sigma =\sqrt{d_1/2(d_1+2)}$ and
$\gamma =\sqrt{2/d_1(d_1+2)}$. We restrict ourselves to the case of a
 potential $V(\chi)$ with a single zero minimum 
 $\left. V(\chi _{min})=V(\chi )\right|_{min} = 0$,
 $\left.\partial ^2_{\chi}V\right|_{min}=m^2_{\chi}>0$
because the non-zero minimum case $\left. V(\chi )
\right|_{min} = V_0$ is trivially reduced to the zero minimum one by
inclusion of $V_0$ into $\Lambda/\kappa^2_0$.

It is easy to check that 
the effective potential \rf{5.1} has a
 global minimum at the point $\varphi =0, \chi = \chi_{min}$
if the bare cosmological constant 
and the scalar curvature
of the internal space are both
negative\footnote{Compact internal spaces with negative
curvature $R_1=-d_1(d_1-1)$ can be constructed, e.g., as hyperbolic
coset manifolds $H^{d_1}/\Gamma ^{d_1}$ (for details see \cite{hyp}).
Here $H^{d_1}$ is
an infinite hyperbolic space and
$\Gamma ^{d_1}$ ---
an appropriate group of discrete isometries.
The coset manifold itself can be imagined to be built up
from a fundamental polyhedron in $H^{d_1}$
with faces pairwise identified.}: $\Lambda <0$, $R_1<0$.
Additionally it is necessary that these parameters are
connected with the compactification scale $b_0$
by a fine-tuning condition
\be{5.3}
2\sigma \Lambda = (\sigma + \gamma )\frac{R_1}{b_0^2}
\quad \Longleftrightarrow
\quad \Lambda = \frac{D-2}{2d_1}\frac{R_1}{b_0^2}\, .
\ee
{}From the 
 Hessian (the mass matrix) of the effective potential
\be{5.1a}
\left(
\begin{array}{cc}
\partial ^2_{\varphi \varphi}&\partial ^2_{\varphi \chi}\\
\partial ^2_{\chi \varphi}&\partial ^2_{\chi \chi}
\end{array}
\right)
\left.U_{eff}\right|_{\varphi =0,\chi _{min}}
=\left(
\begin{array}{cc}
-2\gamma (\sigma +\gamma)\tilde R_1 &0\\
0&m^2_{\chi}
\end{array}
\right)
\ee
it is clear that in contrast with \rf{4.15a} 
the minimum is non-degenerate  and
induces a non-vanishing mass
$m^2_{\varphi}=-2\gamma (\sigma +\gamma )\tilde R_1$
of the modulus field\footnote{These massive modes
(gravitational excitons \cite{GZ(PRD)}) propagate in the
external spacetime and can yield a considerable contribution to the dark matter
in our Universe \cite{GZ(PRD2)}.}.
As result the position $\varphi =0$
is energetically favored and provides the necessary trapping.

The trapping can also be seen directly 
from the equations of motion of the $(\varphi ,\chi )$
system.
Whereas substitution of \rf{5.1}
into the $\chi $ field equation \rf{4.4} does not change this equation,
the equation
\rf{4.3} for the modulus field reads now
\be{5.2}
\ddot \varphi + d_0 \tilde H \dot \varphi = -\frac{\partial U_{eff}}
{\partial \varphi} = - e^{2\sigma \kappa_0 \varphi}\left[ -\frac{1}{\kappa_0}
(\sigma + \gamma )\tilde R_1 e^{2\gamma \kappa_0 \varphi } + \frac
{2\sigma}{\kappa_0}\Lambda +2\sigma V(\chi )\right]\, .
\ee
So, for fine tuned parameters \rf{5.3} the point 
$\varphi =0$, $\chi =\chi_{min}$ is the trivial solution of the system 
\rf{4.4}, \rf{5.2} and  
small linear perturbations around this solution satisfying
\ba{5.4}
\delta \ddot \varphi + d_0 \tilde H \delta \dot \varphi
- 2\gamma (\sigma +\gamma )\tilde R_1 \delta \varphi &=& 0\, ,\nn\\
\delta \ddot \chi + d_0 \tilde H \delta \dot \chi
+ m^2_{\chi} \delta \chi &=& 0
\ea
 are damped by the present friction terms.
The global modulus dynamics is easily understood from the form of the
effective potential. The minimum at $(\varphi =0, \chi =\chi_{min})$
is the deepest point and defines after stabilization of the system
the effective cosmological constant
$\Lambda_{eff} \equiv \left. \kappa^2_0 U_{eff}\right|_{min}
= \tilde R_1/d_1 < 0$
of the external spacetime.
In the small-scale-factor region
we have $U_{eff}(\varphi \to \infty)\to \infty $
and in the
decompactification region
$U_{eff}(\varphi \to -\infty)\to 0$, so that
we are led to the following implications.
1) the Brustein-Steinhardt \cite{BS}
problem\footnote{The Brustein-Steinhardt problem can occur for
potentials with an energetically allowed decompactification region which
is separated
from a narrow valley around the minimum by a barrier.
Supposing that the modulus field rolls down the potential it can overshoot
 this narrow valley and enter the decompactification region. The problem
consists in the fact that the modulus field with high kinetic energy 
will not "feel"
the valley and will not stabilize in the corresponding minimum.}
cannot occur because there is no barrier in the model 
at all, separating the minimum
of the potential from an asymptotically allowed region 
(the single minimum is located
energetically deeper than the decompactification region);
2) higher order approximations to the solution of \rf{5.2} will not alter
the stabilization picture for the same reason; 
and 3) the dying away friction term
$d_0\tilde H\dot \varphi$ will leave at most  small fluctuations of
the geometrical modulus field
around the minimum.
This means, that this minimum
stably compactifies (traps) the internal space. 
Clearly, such a stabilization mechanism 
is different from the above discussed dynamical stabilization
where the effective potential has no  minimum with respect to $\varphi$.

As conclusion of this section we would like to note
that the negative effective cosmological constant
$\Lambda _{eff}<0$
of the model
 \rf{5.1} leads to a turning point in the scale factor 
evolution of the external space, i.e. 
a stage of external space inflation 
is followed be a period of contraction.
This happens when the effective energy density $\rho_{eff}$ of the fields
in the Universe becomes smaller than $|\Lambda _{eff}|$. Model \rf{5.1}
with
stable internal space
 can be brought
in agreement with the observed positive effective cosmological
constant of the Universe, 
e.g. if one takes into account additional higher dimensional
form fields \cite{GZ(nl2)}. 
%%%%%%%%%%%%%%%%%%%%%%%%%%%%%%%%%%%%%%%%%%%%%%%%%%%%%%%%%%
\section{Conclusion\label{conclusion}}
\setcounter{equation}{0}

\bigskip

In the present paper we investigated in detail the
recently proposed \cite{Maz}
 mechanism of  internal space dynamical compactification
in a  multidimensional model with one Ricci-flat internal space and
a massive scalar field acting as inflaton.
As shown
in section \ref{e-frame}, such a dynamical stabilization
could be possible if the modulus field is only coupled to the
inflaton but not to its decay products.
Due to the exponential decay of the inflaton
during reheating the force term in the modulus equation, obtained
from the effective potential, decreases also
exponentially
so that the modulus field can dynamically stabilize via friction.
If the decay products of the inflaton will couple
similar to the modulus field
like the inflaton itself, the dynamical stabilization will not occur.
An analogous decompactification has place for a model with
non-vanishing effective cosmological constant.
 A stable compactification for the considered model can be ensured
via trapping
of the internal space scale factor by a minimum of the effective potential.
An example of such type of stable compactification is given in section
\ref{stable-c}.
Oscillations of the modulus and the inflaton field around the minimum are
observed as massive
scalar particles (gravexitons and inflaton particles) 
in the external spacetime.
Such particles were considered in the electroweak and the planckian
fundamental scale approaches
in Ref. \cite{GZ(PRD2)}. In the same paper it was pointed out
that there exists a rescaling for
gravexciton masses in the different approaches and extremely light particles
may arise in the planckian fundamental scale approach.
Further it was shown there
that particles with masses $m_{\varphi} \sim 10^{-33}$eV are of special
interest because, first, they do not overclose the Universe, second,
the period of their oscillations $T_{osc} \sim 1/m_{\varphi}$ is
of order of the Universe age $\sim 10^{18}$sec, and third,
the effective cosmological term
$\Lambda_{\varphi} \equiv \kappa^2_0 m^2_{\varphi}
\varphi^2 \sim 10^{-57}cm^{-2}$ corresponding to models with such 
 particles and $\varphi \sim M_{Pl}$, takes a value
 of order of the presently observable cosmological constant in the
Universe. Thus, these particles evolve
extremely slowly within their minimum and it is attractive to treat
$\Lambda_{\varphi}$ as an effective cosmological constant
 in the spirit
of quintessence (for models with zero minimum of an effective potential).
However, such a quintessence would have a drawback
because it requires a highly fine tuned initial condition
of the modulus field $\varphi$ (see e.g. \cite{Chicago}). 
%%%%%%%%%%%%%%%%%%%%%%%%%%%%%%%%%%%%%%%%%%%%%%%%%%%%%%%%%%%%%%%%

\bigskip
{\bf Acknowledgments}

We would like to thank Anupam Mazumdar for numerous discussions
and valuable comments, and Maxim Pospelov for drawing our attention to
Refs. \cite{CGKT,KKOP}. 
We also acknowledge useful discussions with Martin Rainer.
The work was finished during A.Z.'s visit at  the University of Minnesota, 
Fermilab and Princeton University.
He thanks Keith Olive and the ITP, 
Edward Kolb and the Fermilab Astrophysics Center, as well as
Paul Steinhardt and the Princeton University for their
kind hospitality.
U.G. acknowledges financial support from DFG grant KON 1575/1999/GU 522/1.

\begin{appendix}
\section{The effective potential
of a higher dimensional perfect fluid\label{appendix}}
In this Appendix we assume that the bulk scalar field $\tilde \chi$ behaves
like a perfect fluid living in the $D-$dimensional bulk spacetime and 
obeying the equation of state
\be{a1}
P_{\tilde \chi}=(\alpha-1)\rho_{\tilde \chi}
\ee
with $0\le \alpha \le 2$ a constant.
The energy density $\rho_{\tilde \chi}\equiv -T^0_0$
of the scalar field $\tilde \chi$ and its pressure
$P_{\tilde \chi}\equiv T^N_N$ are defined from the action functional \rf{2.4}.
Further, we assume that the metric of the $D_0-$dimensional external
spacetime is given by the ansatz \rf{10}
and that a homogeneous approximation for
the scalar fields $\tilde \chi =\tilde \chi (t)$,
$\Phi =\Phi (t)$ can be used.
Then, as it was shown in  
\cite{Z(CQG)}, the action \rf{2.4} of the gravity --- scalar field system
is equivalent to the gravity --- perfect fluid action
\be{a2}
S=\frac{1}{2\kappa ^2_D}\int _M d^Dx\sqrt{|g|}
\left\{R[g]-2\frac{\kappa ^2_D}{v_{d_1}}\rho _{\chi}(a,b)\right\}\, ,
\ee
where
\be{a3}
\rho _{\chi}(a,b)= \frac{A}{a^{\alpha d_0}b^{\alpha d_1}}\, , \qquad
A:=const
\ee
and we took into account the redefinitions \rf{2.8a} of the scalar
field $\tilde \chi \longrightarrow \chi$ and the potential
$U(\tilde \chi) \longrightarrow V(\chi)$.
Dimensional reduction of \rf{a2} gives the Brans-Dicke frame action functional
similar to \rf{1}
\be{a4}
S = \int\limits_{M_0}d^{D_0}x\sqrt{|g^{(0)}|}
\left\{\Phi R\left[ g^{(0)}\right]
-\frac{\omega }{\Phi } g^{(0)\mu\nu }\partial _\mu \Phi \,\partial _\nu
\Phi - 2 \kappa^2_0 \Phi \rho _{\chi (BD)}(a,\Phi )\right\}\, , 
\ee
where the energy density $\rho _{\chi (BD)}$ is defined as
\begin{eqnarray}
\rho _{\chi (BD)}(a,\Phi )
&=&\frac12 \left(\frac{d\chi}{d t}\right)^2+V(\chi)\label{a5}\\
&=&\frac{A}{a^{\alpha d_0}\left(2\kappa_0^2\Phi\right)^{\alpha}
b_0^{\alpha d_1}}\, .\label{a6}
\end{eqnarray}
Finally, via conformal transformation \rf{2.10} we
 pass to
the Einstein frame and arrive at
\be{a7}
S=\frac12 \int_{M_0}d^{D_0}x \sqrt{|\tilde g^{(0)}|}\left\{
\frac{1}{\kappa _0^2}\tilde R\left[\tilde g^{(0)}\right]
-\tilde g^{(0)\mu\nu}\partial_{\mu}\varphi\partial_{\nu}\varphi
-2\rho _{\chi (E)}(\tilde a,\varphi)\right\}\, .
\ee
The energy density $\rho _{\chi (E)}(\tilde a,\varphi)$
 can be recast as
\begin{eqnarray}
\rho _{\chi (E)}(\tilde a,\varphi)
&=&\frac12 \left(\frac{d\chi}{d \tilde t}\right)^2
+e^{2\sigma \kappa_0\varphi}V(\chi)\label{a8}\\
&=&e^{2\sigma \kappa_0\varphi}\rho _{\chi (BD)}(a,\Phi )\label{a9}\\
&=&e^{(2-\alpha )\sigma \kappa_0\varphi}\tilde \rho _{\chi }(\tilde a)\, ,
\label{a10}
\end{eqnarray}
where we used the
relations between the synchronous times
$d\tilde t=\pm e^{-\sigma \kappa_0\varphi}dt$
and the scale factors
$\tilde a=e^{-\sigma \kappa_0\varphi}a$
in the Brans-Dicke frame $(t,a)$
and the Einstein frame $(\tilde t, \tilde a)$
and defined
the reduced energy density $\tilde \rho _{\chi}$  as
\be{a11}
\tilde \rho _{\chi}(\tilde a)
\equiv \frac{A}{\tilde a^{\alpha d_0}b_0^{\alpha d_1}}\, .
\ee

The equations of motion following from action functional \rf{a7} for
homogeneous field $\varphi =\varphi (\tilde t)$ and energy density
$ \rho _{\chi (E)}=  \rho _{\chi (E)}(\tilde t)$,
and Ricci-flat external space
$R[\bar g^{(0)}]=0$ read:
\begin{eqnarray}
\tilde H^2\equiv \left(\frac{\dot {\tilde a}}{\tilde a}\right)^2
&=&\frac{2\kappa _0^2}{d_0(d_0-1)}\left\{\frac12\dot \varphi ^2+
\rho _{\chi (E)}(\tilde a,\varphi)\right\}\, ,
\label{a12}\\
\ddot \varphi +d_0\tilde H\dot \varphi
&=&(\alpha -2)\sigma\kappa _0\rho _{\chi (E)}(\tilde a,\varphi)\, ,
\label{a13}\\
\dot {\tilde H}
&=&-\frac{\kappa _0^2}{d_0-1}\left\{\dot \varphi ^2+
\alpha \rho _{\chi (E)}(\tilde a,\varphi)\right\}
\label{a14}\, .
\end{eqnarray}
Eqs. \rf{a10} and \rf{a13} show that for $\alpha =2$ the interaction
between the modulus field $\varphi$
and the perfect fluid (scalar field $\chi$)
is absent. This fact is easily explained because $\alpha =2$ corresponds to
a perfect fluid with ultra-stiff equation of state, which describes a scalar
field with vanishing potential energy $V(\chi)\equiv 0$.
The other extremal case with $\alpha =0$ describes a vacuum equation of state
and can be used for  potentials $V(\chi)$ with non-zero minimum $V_0$.

At the end of this Appendix let us consider a model 
with stabilizing modulus field $\varphi \approx 0$ and decaying field $\chi$.
Similar to model (II) of section \ref{e-frame}
we assume that the decay products with energy density
$\rho _{SM (E)}\sim \tilde a^{-\alpha d_0}$
are not coupled to the modulus field, but that they, nevertheless,
define the dynamics
of the external space.
Then the corresponding equation system follows from \rf{a12}, \rf{a14}
and reads
\begin{eqnarray}
\tilde H^2
&=&\frac{2\kappa _0^2}{d_0(d_0-1)}
\rho _{SM (E)}(\tilde a)\, ,
\label{a15}\\
\dot {\tilde H}
&=&-\frac{\alpha \kappa _0^2}{d_0-1}
 \rho _{SM (E)}(\tilde a)\, ,
\label{a16}
\end{eqnarray}
so that the Hubble parameter behaves as $\tilde H=s/\tilde t$ with
\be{a17}
s=\frac{2}{d_0\alpha}\, .
\ee

\end{appendix}

%%%%%%%%%%%%%%%%%%%%%%%%%%%%%%%%%%%%%%%%%%%%%%%%%%%%%%%%%%%%%%%%%
%%
%%

%
%%%%%%%%%%%%%%%%%%%%%%%%%%%%%%%%%%%%%%%%%%%%
\end{document}